\documentclass{IEEEtran}
\usepackage{cite}
\usepackage{amsmath,amssymb,amsfonts}
\usepackage{algorithmic}
\usepackage{graphicx}
\usepackage{textcomp}
\usepackage{subfigure}
\usepackage{color}
\usepackage{hyperref}
\usepackage{bm}

\newcommand{\f}{\varphi}
\newcommand{\w}{\omega}

\def\BibTeX{{\rm B\kern-.05em{\sc i\kern-.025em b}\kern-.08em
    T\kern-.1667em\lower.7ex\hbox{E}\kern-.125emX}}
\begin{document}
\title{Metasurface-Enabled Interference Suppression\\ at Visible-Light Communications}
\author{Constantinos Valagiannopoulos$^*$, Theodoros A. Tsiftsis and Vassilios Kovanis

\thanks{This work has been supported by Nazarbayev University Faculty Development Competitive Research Grant No. 090118FD5349. Funding from MES RK state-targeted program BR05236454 is also acknowledged.}
\thanks{$^*$Constantinos Valagiannopoulos is with Nazarbayev University, KZ-010000, Kazakhstan and the corresponding author (Email: {\tt konstantinos.valagiannopoulos@nu.edu.kz}).}
\thanks{Theodoros Tsiftsis          is with Jinan University, CN-519070, China.}
\thanks{Vassilios Kovanis              is with Nazarbayev University, KZ-010000, Kazakhstan and Virginia Tech Research Center, VA-22203, USA.}}

\maketitle

\begin{abstract}
Light can be used for wireless information transmission apart from illumination; that is the key idea behind visible-light communication (VLC), one of the disruptive technologies of our days. It combines remarkably high data rates due to ultrashort wavelengths with huge reliability and security due to small distances; nevertheless, it substantially suffers from interference of neighboring light sources in multiple-link configurations. In this manuscript, we investigate a pair of light emitting diodes (LEDs) interfering each other and propose a simple nanoslit metasurface that radically increases the directivity of the transmitting beams. As a result, enhancement of the signal-to-interference ratio by several orders of magnitude is reported. The considered generic setup retains its beneficial features in the presence of realistic design defects and, accordingly, may inspire standardization efforts towards the adoption of VLC in next-generation heterogeneous communication networks.
\end{abstract}


\section{Introduction}
\label{sec:introduction}
The visible part of the spectrum occupies over 600 THz of license-free band that offers a unique possibility for design, testing and implementation of wireless links supporting huge data rates. The basic idea \cite{FirstPaper} is to exploit the existing infrastructure of the lighting system to modulate and encode the information signal on top of the illumination \cite{BookDimitrovHaas, BookArnon}. Visible-light communications (VLC) has recently attracted substantial scientific, funding and industrial interest due to the extensive use of high power light emitting diodes (LEDs) in the visible spectrum \cite{VLCStateOfArt} and the implementation of well-established physical and medium-access layer technologies \cite{VLCChallenges}. 

VLC major advantages pertain to accomplishing dense and high-speed wireless communication in indoor environments such as aircraft cabins, retail shops, professional offices or private homes, combined with reduced power consumption and improved robustness \cite{HighSpeedVLC}. Most importantly, one can deploy a large-scale network supporting several parallel links with minimal infrastructure, standardization and coordination efforts, since visible band is vacant of communication signals \cite{VLCPoint2Point}. Finally, given the fact that light can be contained in a physical space, top security levels are reached for VLC just by shutting doors and windows \cite{VLCSecurity}. Possessing all these values, it makes no wonder that VLC are becoming widespread, make their way towards commercialization \cite{VLCMarket} and offer a sufficient incentive for potential companies \cite{PureLiFi}.    

As stated above, VLC work with multiple LEDs, placed close one-another, to achieve massive parallelization of data streams; therefore, interference between the transmitting beams of neighboring LEDs is the main bottleneck in the VLC operation. For this reason, various elaborate studies have been performed in order to suppress the harming effect of interference with help from power control and orthogonal propagation \cite{InterferenceManagement} or by suitably changing the signal amplitude in time \cite{DemonstrationBiDirectional}. Moreover, the intervention between photodiodes receiving at neighboring frequencies have been mitigated via deploying a spectrum sensor array at the receiver side, assisted by robust signal fusion algorithms \cite{InterferenceRejection}. It should be stressed that, apart from inter-LED interference, ambient light or reflections from various indoor objects can significantly degrade VLC efficiency; accordingly, the necessary estimation of the optical wireless channel has been performed by taking into account multipath dispersion characteristics \cite{IndoorChannel} or mixed specular-diffuse reflections \cite{ChannelModeling}.

With the advent of metasurfaces \cite{SmithReview}, the two-dimensional sibling of metamaterials, the manipulation and transformation of emitted beams became feasible within ultrashort distances. It is thus natural that metasurfaces are being extensively employed to exotic beamforming applications like developing three-dimensional cusp beams for super-resolution imaging \cite{Generation3DCusp} or demultiplexing spontaneous emission signals along different directions \cite{OptoVLSIBeamformer} and with different angular momentum order \cite{OrbitalAngularMomentum}. However, even the simplest form of metasurface, a metallic nanoslit array \cite{AnalyticTheory}, remains popular and very efficient when controlling the transmitting beams. By utilizing easy-to-fabricate designs, one can accomplish optical switching \cite{OpticalSwitching}, subwavelength focusing \cite{ActiveControl} and enhanced Fano resonances \cite{NanoslitArrays} with visible light.

In this work, we propose a similar nanoslit array to remedy the aforementioned interference and reflection problems occurred at closely placed LEDs providing multiple parallel VLC links. Such a primitive metasurface, placed on top of the Lambertian transmitter, boosts dramatically the directivity of the emitted radiation pattern and suppresses accordingly the intercessions between neighboring receivers; importantly, these trends are increasing with the number of slits. Based on rigorous mathematical formulations for the electromagnetic fields, we analytically derive the signal-to-interference ratio (${\rm SIR}$) and the received power of the signal through Poynting theorem \cite{AntennaBalanis}. By comparing the quality of the link with and without the metasurface, we report enhancement of the ${\rm SIR}$ by several orders of magnitude accompanied only by a moderate decrease of the received power, enabling higher data rates. Since with the proposed configuration, the vast portion of emitted power is concentrated across a tiny cone, the system becomes vulnerable to a (translationally or rotationally) misaligned receiver and pointing errors. However, we find that, with realistic sizes and angles, the described setup performance exhibits significant insensitivity with respect to design defects. The proposed concept combines structural simplicity and increased throughput with substantial robustness, which can pave the way for 5G light fidelity (Li-Fi) standards.

The remainder of this paper is structured as follows. In Section \ref{sec:system}, the system model is presented, the electric field response is exactly determined, the radiation pattern of the metasurface is found and certain power considerations are made. In Section \ref{sec:numres}, we define the ranges for the input parameters, investigate the beneficial influence of the metasurface on interference suppression and test the sensitivity of the setup to a misplaced receiving photodiode. Finally, in Section \ref{sec:conclusion}, concluding remarks are delivered and aspects for the perspective of VLC market are outlined.

\section{System Model}
\label{sec:system}

\subsection{Electric Field Response}
\begin{figure}[ht!]
\centering
\includegraphics[width=7.9cm, draft=false]{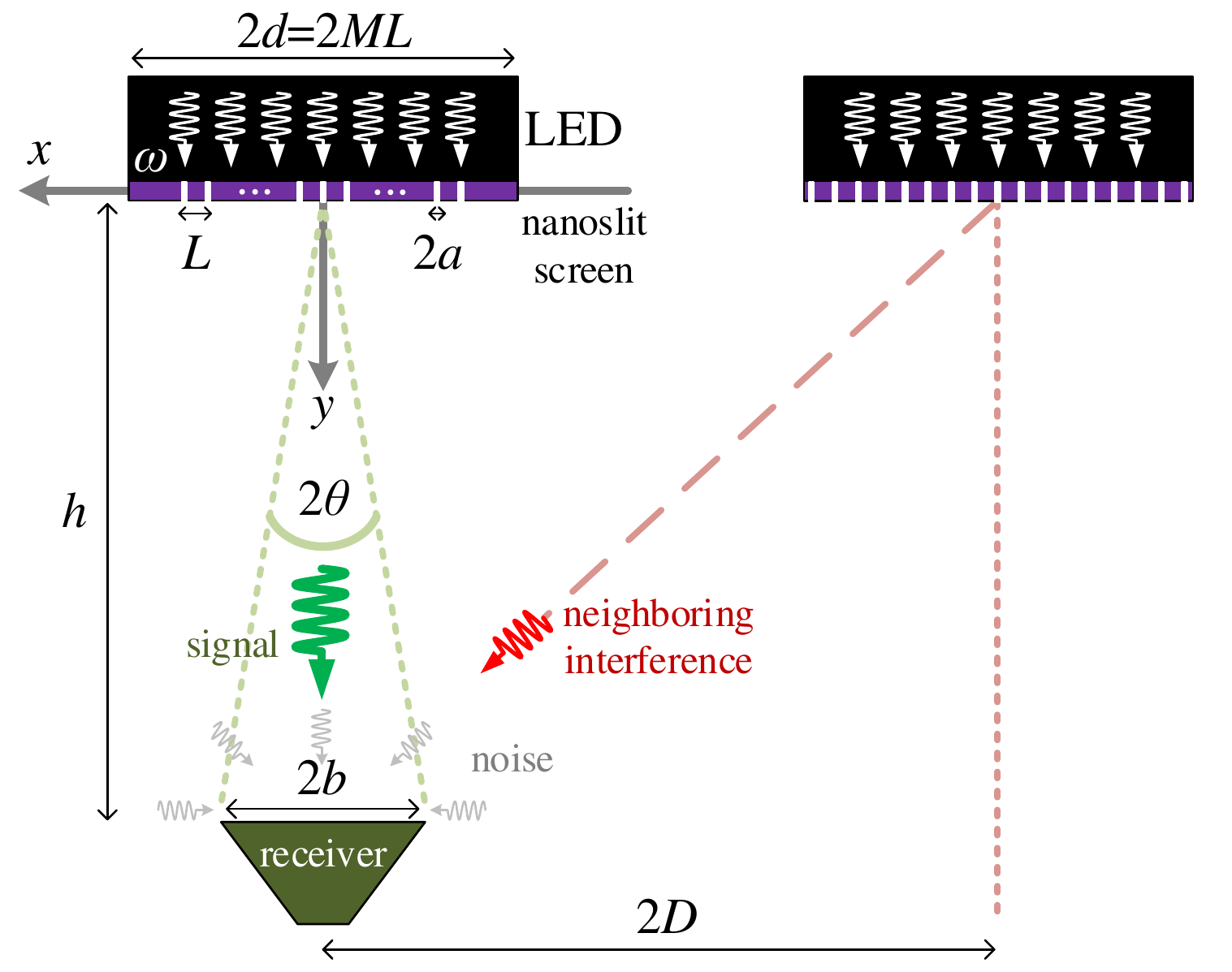}
\caption{Simplistic two-dimensional sketch of visible-light communication between modulated LED signals passing through properly patterned metasurface of multiple nanoslits and distant receiver, which gets interfered by a neighboring identical LED source.}
\label{fig:Fig1}
\end{figure}

We assume that one has properly modulated the light from a LED source to carry information signals within an indoor environment. In Fig. \ref{fig:Fig1}, where the used Cartesian coordinate system $(x,y,z)$ is also defined, we show the sketch of a configuration supporting communication between the aforementioned source and the receiver. The light from LED passes through a patterned metasurface comprising an array of very small slits of size $2a$ crafted on a metallic impenetrable plane. It is supposed that the reflections are handled properly by a matched absorber behind LED without creating standing waves and bothering the data stream emission. The slits are $(2M+1)$ in number and very lengthy along $z$ axis; in addition, they are equispaced with distance $L$ between (the centers of) two consecutive ones. Without loss of generality, we can take electric field to be polarized along $z$ direction. Therefore, the formulation becomes two-dimensional with (Fourier transform of) scalar field $E_z(x,y)$, since the structure is practically invariant with respect to $z$. 

Note that the LED source emits light across the entire visible frequency $\w$ range and thus it can modulate multiple signals around several central frequencies $\w_0$. As far as the other end of the communication link is concerned, we consider a single photodiode receiver with an aperture of $2b$, positioned at distance $h$ along the normal direction of the radiating array. Again, we do not elaborate polarization concerns regarding the orientation of the VLC user since the size along $z$ axis is very large. The length of the emitter is denoted by $2d=2ML$, while an identical one positioned at horizontal (along $x$ axis) distance $2D$ interferes the communication. The followed approach is rigorous and can be trivially generalized to treat multiple LED/photodiode pairs; thus, it is fundamentally different than studies with similar objectives treating channel uncertainty through coordinated beamforming \cite{CoordinatedBeamforming} or optimizing the position of receivers for different radiation patterns \cite{ImpactLEDTransmitters}. In the present analysis, harmonic time of the form $e^{+j\w t}$ is suppressed for the field quantities $E_z$, namely the Fourier transform $S(\w)$ of the time signal $s(t)$ is defined as: $S(\w)=\int_{-\infty}^{+\infty}s(t)e^{-j\w t}dt$.

The size $2a$ of the slits and the length of the LED source $2d$ can be considered as inherent characteristics of the transmitting system. Indeed, in a realistic context, a cluster of multiple identical emitters will be paired with the respective VLC users positioned on a planar rectangular panel giving high parallelization in information transmission. Therefore, the number $N=\lceil \frac{d-a}{2a} \rceil$, which corresponds to the $(2N+1)$ nanoslits that fit within the radiating aperture if each one of them is placed next to the other ($L=2a$), makes a specification of the propagation system with $N \ge M \ge 1$.

Due to the small size of the slits along $x$ axis, we can assume with no loss of generality that each of them can work as a line source (thin cylinder of radius $a$) radiating omnidirectionally (for the sake of brevity) into the vacuum space $y>0$. The intensity of each source is the same since the LED field is taken as homogeneous and the complex (electric) fields developed at $x=mL$ for $m=-M,\cdots,M$ take unitary value (1 $Volt/meter$). In this way, the collective response (diffused $z$-directed electric field) from all the $(2M+1)$ nanoslits for $y>0$, takes the form \cite{CylWaves}:
\begin{eqnarray}
E_z(x,y)=\sum_{m=-M}^M \frac{H_0\left(\frac{\w}{c}\sqrt{(x-mL)^2+y^2}\right)}{H_0(\w a/c)},
\label{OutputField}
\end{eqnarray}
where $H_0$ is the Hankel function of zero order and second type and $c$ is the speed of light into vacuum. The normalization constant $H_0(\w a/c)$ in the denominator of \eqref{OutputField} is placed to give unitary field in the vicinity of each nanoslit. 

In the far region, which is the vast part of an office room for visible light frequencies (distance $h$ from metasurface), where inevitably the receiver is placed, the Fourier transform of the electric field is given by \cite{Abramowitz}:
\begin{eqnarray}
E_z(r,\f,\w)\cong \sqrt{\frac{2cj}{\pi\w r}}\frac{e^{-j\frac{\w}{c}r}}{H_0(\w a/c)}\sum_{m=-M}^M e^{j\frac{\w}{c} L m \cos\f},
\label{OutputFarField}
\end{eqnarray}
expressed in the corresponding polar coordinates $(r,\f)$, where $\f=0$ along positive $x$ semi-axis. The actual electric field in time domain can be evaluated by taking the inverse Fourier transform as: 
\begin{eqnarray}
e_z(r,\f,t)=\frac{1}{2\pi}\int_{-\infty}^{+\infty}Q(\w)E_z(r,\f,\w)e^{j\w t}d\w,
\label{TimeElectricField}
\end{eqnarray}
 where $Q(\w)$ is the Fourier transform (measured in $seconds$) of the pulses $q(t)$ comprising the information signal \cite{Pulses} modulated around $\w=\w_0$. In the simplest scenario, one may consider a single tone $q(t)=e^{j\w_0 t}$ with $Q(\w)=2\pi \delta(\w-\w_0)$ and thus, due to linearity, $e_z(r,\f,t)=E_z(r,\f,\w_0)e^{j\w_0 t}$. In the following, we will confine ourselves to such harmonic excitations since our primary goal is to examine the effect of the metasurface structural parameters on the characteristics of the emitted beam.

\subsection{Metasurface Radiation Pattern}
The radiation pattern $G(\f)$ of the device, expressing the intensity developed far from the metasurface, if the cylindrical wave factor $\sqrt{\frac{2cj}{\pi\w_0 r}}e^{-j\frac{\w_0}{c}r}$ gets dropped in (\ref{OutputFarField}), is written as \cite{AntennaBalanis}:
\begin{eqnarray}
G(\f)=\frac{1}{|H_0(k_0a)|^2}\left[\frac{\sin\left(\frac{k_0L}{2}(2M+1)\cos\f\right)}
                 {\sin\left(\frac{k_0L}{2}\cos\f\right)}\right]^2,
\label{RadiationPattern}
\end{eqnarray}
where the geometrical progression sum in \eqref{OutputFarField} has been analytically evaluated. The numerator is nullified at $\cos\f=\frac{\lambda}{L}\frac{v}{2M+1}$ for integer $v\in\mathbb{Z}$, where $\lambda=2\pi c/\w_0=2\pi/k_0$ is the oscillation wavelength into free-space ($k_0=\w_0/c=2\pi/\lambda$ is the corresponding wavenumber of the central working frequency $\w_0$). Since the denominator becomes also zero for $\cos\f=\frac{\lambda}{L}w$ for integer $w\in\mathbb{Z}$, which means that neutralizes one zero in $(2M+1)$ ones of the numerator, the total radiation pattern gets nullified approximately:
\begin{eqnarray}
V\cong \left\lfloor 2 \left\lfloor (2M+1)\frac{L}{\lambda}\right\rfloor \frac{2M}{2M+1} \right\rfloor
\cong  \left\lfloor \frac{4d}{\lambda} \right\rfloor
\label{LobesNumber}
\end{eqnarray}
times. $V$ corresponds also to the number of maxima of the radiation pattern (in-between two consecutive zeros), namely the lobes of the produced beam. 

The largest value of $G(\f)$ is taken at $\f=\pi/2$ (due to the pole-zero cancellation for $v=w=0$) and equals to $G(\pi/2)=\frac{(2M+1)^2}{|H_0(k_0a)|^2}$. Obviously, such a maximal value cannot increase unboundedly since the nanoslits are of finite size $2a$; for this reason, its upper bound is exhibited for $M=N$, namely the nanoslits cover the entire radiating aperture without blocking any portion of the incoming illumination from LED ($L=2a$). One evaluates the flux corresponding to the intensity profile of \eqref{RadiationPattern} for that scenario ($M=N$ and accordingly $L=2a$) as follows \cite{FluxIntensity}:
\begin{eqnarray}
g_{max}\cong\frac{1}{|H_0(k_0a)|^2}\int_0^{\pi}\!\frac{\sin^2\left(k_0d\cos\f\right)}
                                                    {\sin^2\left(k_0a\cos\f\right)}d\f.
\label{MaximumFlux}
\end{eqnarray}
Therefore, it is natural to assume that the radiation pattern of LED chip itself (without metasurface) is given by:
\begin{eqnarray}
G_0(\f)=\frac{g_{max}}{2}\sin\f,
\label{RadiationPattern0}
\end{eqnarray}
namely the typical Lambertian emission \cite{LambertianPattern} from a homogeneous flat surface with the same flux $g_{max}$.

\begin{figure}[ht!]
\centering
\subfigure[]{\includegraphics[width=6.0cm, draft=false]{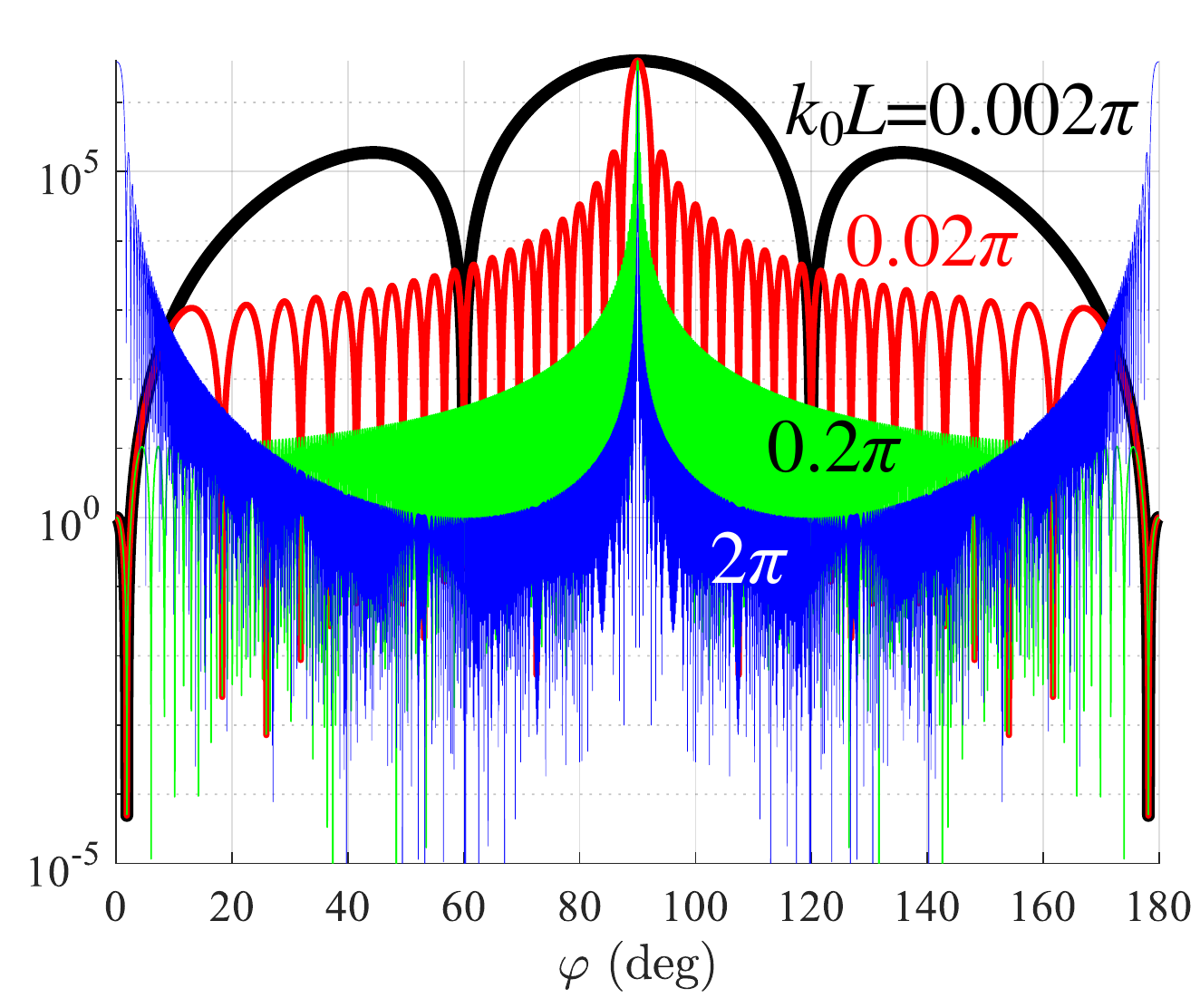}
   \label{fig:Fig2a}}
\subfigure[]{\includegraphics[width=6.0cm, draft=false]{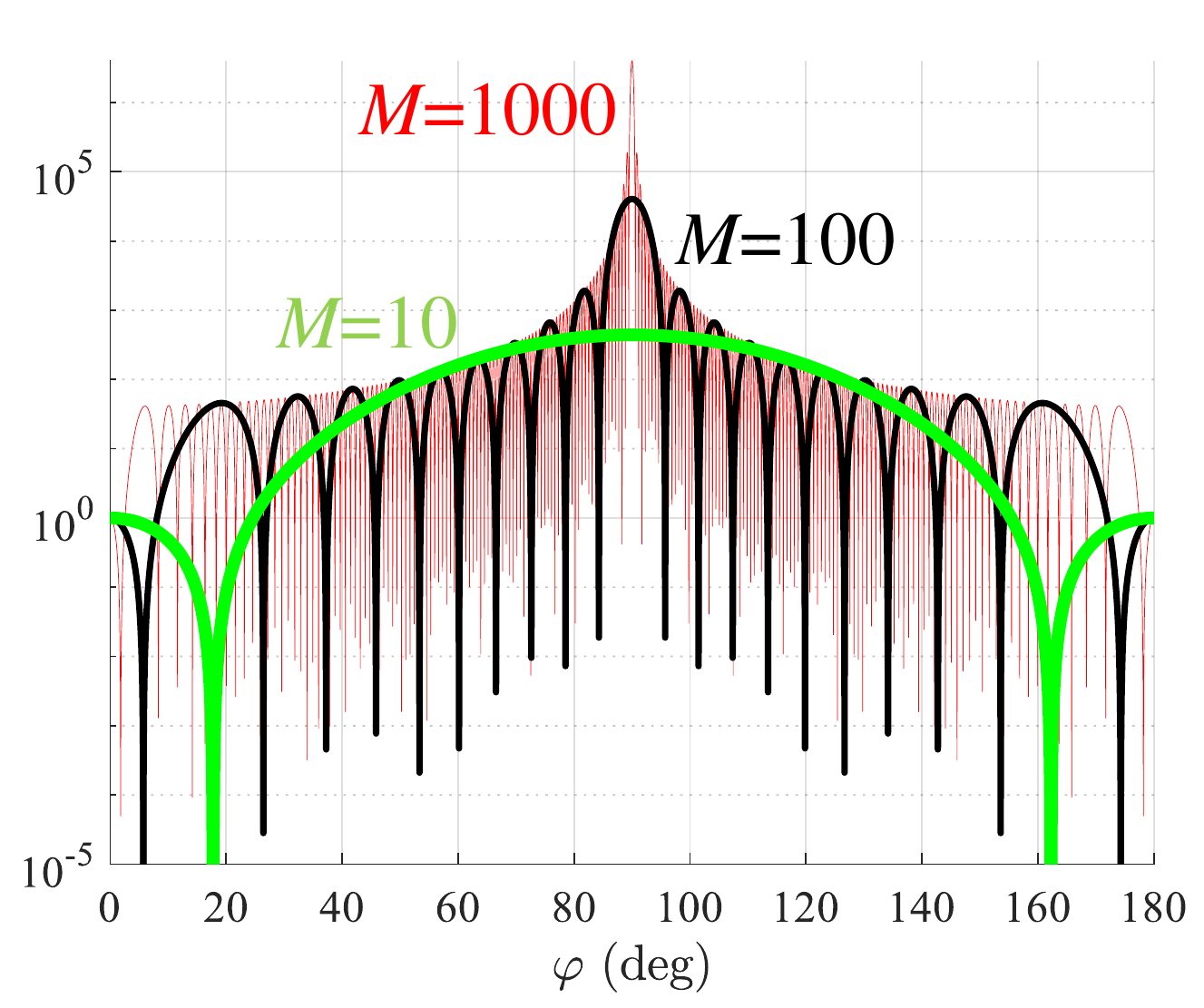}
   \label{fig:Fig2b}}
\caption{The radiation patterns $G(\f)|H_0(k_0a)|^2$ as functions of azimuthal angle $\f$ for: (a) several metasurface optical periods $k_0L$ ($M=1000$) and (b) several numbers of nanoslits $M$ ($k_0L=0.1\pi$).}
\label{fig:Figs2}
\end{figure}

In Fig. \ref{fig:Fig2a}, we show the variation of $G(\f)$ (multiplied by the constant $|H_0(k_0a)|^2$) across half space $y>0$ for a metasurface with fixed large ($M=1000$) number of nanoslits and various electrical spatial periods $k_0L$. Since usually $d\gg L \Rightarrow M\gg 1$, the function $G(\f)$ possesses a rapidly varying numerator and a slowly varying denominator. As indicated above, the maximum value (major lobe) is exhibited at $\f=\pi/2$ and is equal to $(2M+1)^2$ regardless of $k_0L$. In addition, the larger is the optical distance $k_0L$, the more rapid is the azimuthal variation of $G(\f)$ since the frequency of the numerator's harmonic function gets increased (the number $V$ of lobes given by \eqref{LobesNumber} is proportional to $k_0L$). Notice also that, unlike main lobe, the sidelobes see their level getting shrunk for larger $k_0L$; such a behavior is attributed to the proportional growth of the denominator of \eqref{RadiationPattern} far from maximal direction, which makes the fraction smaller (exactly at $\f=\pi/2$, we have pole-zero cancellation). By inspection of Fig. \ref{fig:Fig2a}, it is clear that the directivity of the beam, which offers the isolation of the emitter/receiver pair from neighboring systems, gets enhanced once $k_0L$ increases. However, since the denominator of $G(\f)$ is also a periodic function, such a trend has its limit; indeed for $k_0L>2\pi$, new strong lobes emerge and interfere substantially. For this reason, a good choice can be $k_0L=\pi$ which combines very directive major lobe, since its angular extent is given by:
\begin{eqnarray}
2\theta=2\arcsin\left(\frac{2\pi}{k_0L(2M+1)}\right),
\label{DirectiveAngularExtent}
\end{eqnarray}
with significantly suppressed sidelobes.

In Fig. \ref{fig:Fig2b}, we represent the same quantity as in Fig. \ref{fig:Fig2a} with fixed metasurface period $k_0L$ this time but for several numbers $M$ of nanoslits. Obviously, the maximum of the main lobe at $\f=\pi/2$ increases like $(2M+1)^2$ as mentioned above; the sidelobes get also larger but more reluctantly. In particular, it can be easily found that the amplitude of the second strongest lobe appeared at $\f\cong \arccos\left(\frac{3\pi}{k_0L(2M+1)}\right)$, equals to $\csc\left(\frac{3\pi}{2(2M+1)}\right)^2$, an increasing function of $M$. However, the major lobe amplitude increases more rapidly and can be up to $(9\pi^2/4)$ times larger than the secondary sidelobe (in the limit $M\rightarrow +\infty$). It is also apparent from Fig. \ref{fig:Fig2b} that all the lobes (main and side) become more directive for higher $M$, as happens for greater $k_0L$ in Fig. \ref{fig:Fig2a}.

This exceptional directivity of the major lobe is related to the aperture of the receiving photodiode. Indeed, the receiver should be wide enough to absorb at least the illumination of angular extent $2\theta$ of the main direction from \eqref{DirectiveAngularExtent}. In particular, $\tan\theta<b/h$ yielding to a minimum number of nanoslits: $(2M+1)>\frac{2\pi}{k_0L}\frac{h}{d}$, which is a large limit since usually $h\gg d$ (and $h\gg b)$. This scenario is schematically depicted in Fig. \ref{fig:Fig1}, where $2\theta$ angular sector just covers the photodiode size $2b$. Interestingly, we conclude that for moderate values of $k_0L$ close to $k_0L=\pi$ (as dictated by directivity objective), the radiating aperture is electrically large: $k_0d\gg 1$, which is the case of LEDs.

\subsection{Power Considerations}
Since we have exact knowledge of the developed electromagnetic field, we can compute exactly the power received by the VLC user, both because of the primary LED and the interfering one positioned in its spatial vicinity. Specifically, the Poynting vector \cite{AntennaBalanis}, expressing the density of the average power flow in the steady state, is radial and proportional to the radiation pattern \eqref{RadiationPattern}: $\textbf{p}(r,\f)=\hat{\textbf{r}}\frac{G(\f)}{\eta_0 \pi k_0 r}$ (multiplied by 1 $(Volt/meter)^2$, where $\eta_0$ is the wave impedance into free space). By integrating only that component of \textbf{p} which is normal to the photodiode aperture ($y$), namely $\textbf{p}\cdot\hat{\textbf{y}}=\frac{G(\f)}{\eta_0 \pi k_0 r}\sin\f$, one can compute the signal-to-interference-ratio in the considered setup of Fig. \ref{fig:Fig1} as follows:    
\begin{eqnarray}
{\rm SIR}=2\frac{\int_0^{b}        \frac{G(\arctan(h/x))}{x^2+h^2}dx}
                {\int_{2D-b}^{2D+b}\frac{G(\arctan(h/x))}{x^2+h^2}dx}.
\label{SIRFormula}
\end{eqnarray}
Similarly, the signal-to-interference-ratio ${\rm SIR_0}$ in the absence of the metasurface can be found by incorporating $G_0(\f)$ from \eqref{RadiationPattern0}:
\begin{eqnarray}
{\rm SIR_0}=\frac{2b}{\sqrt{h^2+b^2}
\left(\frac{2D+b}{\sqrt{h^2+(2D+b)^2}}-\frac{2D-b}{\sqrt{h^2+(2D-b)^2}}\right)}.
\label{SIRFormula0}
\end{eqnarray}

An important quantity which will be useful as reference power since it is only dependent on the LED size $2d$, the  nanoslit radius $a$ and the operational frequency $\w$, is the total emitted power of the LED in the absence of the metasurface and thus without any reflections from the screen. It is again determined with help from the Poynting vector combined with \eqref{RadiationPattern0} and given by: $P_{max}=\frac{g_{max}}{\eta_0 \pi k_0}$, multiplied by 1 $(Volt/meter)^2$. It should be stressed that $P_{max}$ takes into account all the radiation directions of the Lambertian emission by the LED; on the contrary, the VLC user's aperture covers only a small angular extent from the input wavefront, given that $b\ll h$. For this reason, we do not expect the power of the received signal to be close to $P_{max}$, which will be mainly utilized for normalization purposes. Note finally that all the power quantities are measured in $Watt/meter$ since our structure is infinite along $z$ axis and the corresponding energy flows are defined per unit length (of $z$ axis).

\begin{figure}[ht!]
\centering
\subfigure[]{\includegraphics[width=6.0cm, draft=false]{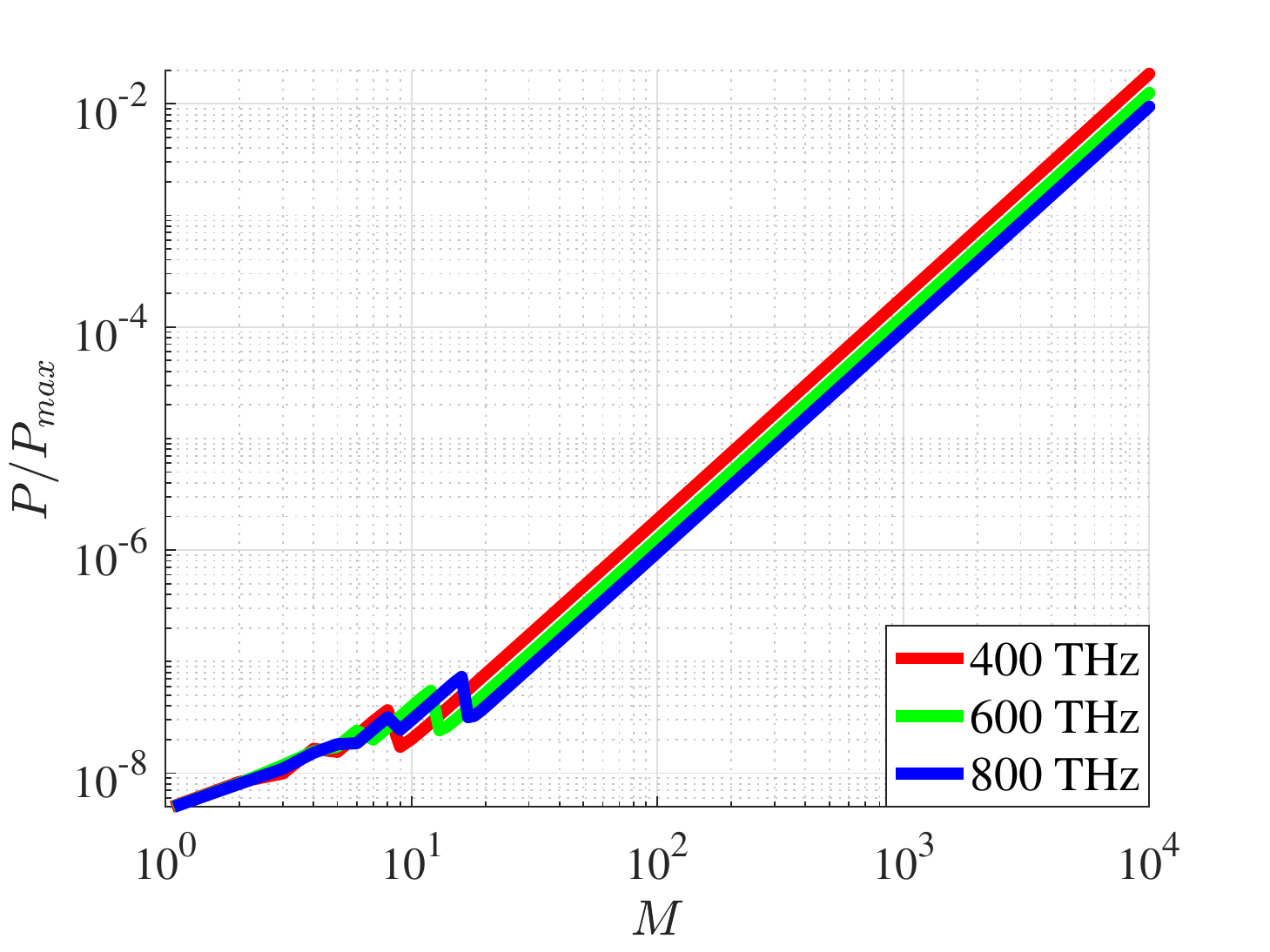}
   \label{fig:Fig3a}}
\subfigure[]{\includegraphics[width=6.0cm, draft=false]{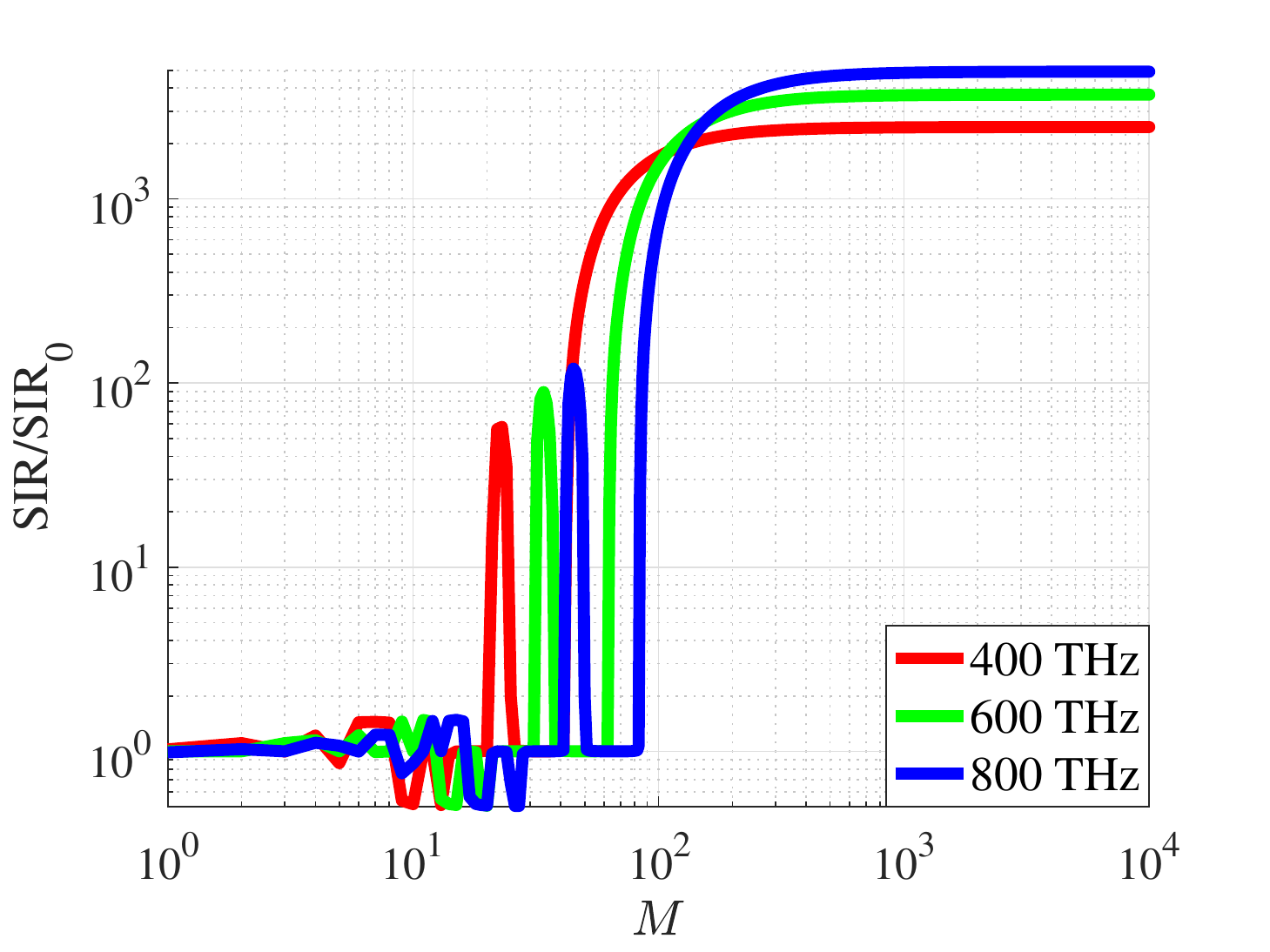}
   \label{fig:Fig3b}}
\caption{(a) Received power of the signal $P$ normalized with the total emitted power in the absence of metasurface $P_{max}$. (b) Signal-to-interference enhancement due to the metasurface ${\rm SIR/SIR_0}$. The quantities are represented as functions of the number of nanoslits $M$ for several operational frequencies $\w/(2\pi)$. Plot parameters: $d=2.5~{\rm mm}$, $a=25~{\rm nm}$, $h=1000~{\rm mm}$, $b=D/2=d$.}
\label{fig:Figs3}
\end{figure}

\section{Numerical Results}
\label{sec:numres}

\subsection{Input Parameters}
It is meaningful to determine the lower and upper limits of the parameters defining our model starting from the operational frequency which covers the entire visible spectrum: $(2\pi)400~{\rm THz}<\w~<(2\pi)800~{\rm THz}$. As far as the LED is concerned, we will keep fixed aperture size $2d=5~{\rm mm}$ \cite{ChannelModeling} and, similarly, the diameter of nanoslits is also kept constant: $2a=50~{\rm nm}$ (maximum number of nanoslits $N\cong 10^5$). The effect of the distance $h$ between emitter and the photodiode is not crucial and thus we assume $h=1000~{\rm mm}$ \cite{ImpactLEDTransmitters}. The receiving aperture should be kept smaller or equal of the LED diameter, namely $b\le d$, so that confined vertical VLC is achieved and, accordingly, the horizontal distance of the interfering LED respects $D\ge d$. The number $M$ of nanoslits can vary from a moderate integer to many thousands (but always $M\ll N$). 

\subsection{Metasurface Effect}
It is crucial to evaluate both the received power (an absolute metric) and the improvement in the signal-to-interference ratio when the metasurface is employed (a relative metric). In Fig. \ref{fig:Fig3a}, we represent the power at the receiver $P$ divided by the total developed power when the LED emits alone $P_{max}$ as function of the number of nanoslits $M$ for three characteristic operational frequencies $\w/(2\pi)$ corresponding approximately to three basic colors of visible light (red, green, blue). Remember that $P_{max}$ is the fixed (for specific $d$, $a$ and $\w$) total emitted power towards all directions in the absence of metasurface, not the received power by the aperture of the VLC user (to be denoted as $P_0$). In other words, it is a very large quantity which is higher than the emitted power in the presence of the nanoslits since no reflections are taken into account; thus, the ratio $P/P_{max}$ is natural to possess small values (especially since $M\ll N$), as happening in Fig. \ref{fig:Fig3a}. 

In particular, we notice a clear increasing trend of all the three curves with $M$ due to two main reasons. Firstly, because of the enhanced maximal power of the produced beam as shown in Fig. \ref{fig:Fig2b}, meaning that a receiver of specific size $b$ collects more energy; secondly, because more and more openings are created in the impenetrable screen at the LED admitting larger portion of the total power $P_{max}$ to exit from it. Surprisingly, the larger is the working frequency, the smaller is the received power; by amplifying $\w=k_0c$, the electrical period $k_0L$ of the metasurface increases which makes the formed radiation pattern more directive but with the same peak (see Fig. \ref{fig:Fig2a}) and, accordingly, less power developed along the fixed length $b$ of the photodiode.

In Fig. \ref{fig:Fig3b}, we depict the variation of the signal-to-interference ratio enhancement ${\rm SIR/SIR_0}$ with respect to $M$ for the same operational frequencies as in Fig. \ref{fig:Fig3a}. One directly notices the abrupt changes in the represented quantity when $M$ is kept moderate ($M \lesssim 50$) which are also present in Fig. \ref{fig:Fig3a} and attributed to the interplay between slits emerging when the number of perforations is small, namely the homogenization of metasurface infeasible \cite{Judicious}. The sharp peaks occur when $k_0L=\frac{d}{c}\frac{\w}{M}$ takes a specific value igniting resonance and that is why they appear at larger $M$ for more substantial frequencies $\w$. However, for $M \gtrsim 300$, a kind of saturation threshold, thousand-fold improvement in signal-to-interference ratio is recorded. Such huge enhancement is due to the corresponding boost of LED beam directivity demonstrated by Fig. \ref{fig:Fig2b}, which secures higher received power (as also indicated by Fig. \ref{fig:Fig3a}) and simultaneously suppresses the interference from the neighboring emitter. 

The aforementioned dramatic increase in ${\rm SIR}$, directly implies that the communication link not only becomes immune to the intervention from other emitters but also mitigates reflections from the surroundings since no diffracted beams are developed. As far as the effect of the frequency is concerned, it is the opposite compared to that of Fig. \ref{fig:Fig3a} since we now evaluate the relative increase in ${\rm SIR}$ which is favored by the directivity of the pattern and not by the power it transmits. Consequently, higher $\w=k_0c$ gives greater $k_0L$ which, in turn, enhance the angular selectivity of the major lobe as shown in Fig. \ref{fig:Fig2a}. If one takes into account Figs. \ref{fig:Fig3a} and \ref{fig:Fig3b} in a combined way, we can understand that a few hundreds of nanoslits provide both very high interference enhancement ${\rm SIR/SIR_0}$ and quite large received power $P/P_{max}$. 

\begin{figure}[ht!]
\centering
\subfigure[]{\includegraphics[width=6.0cm, draft=false]{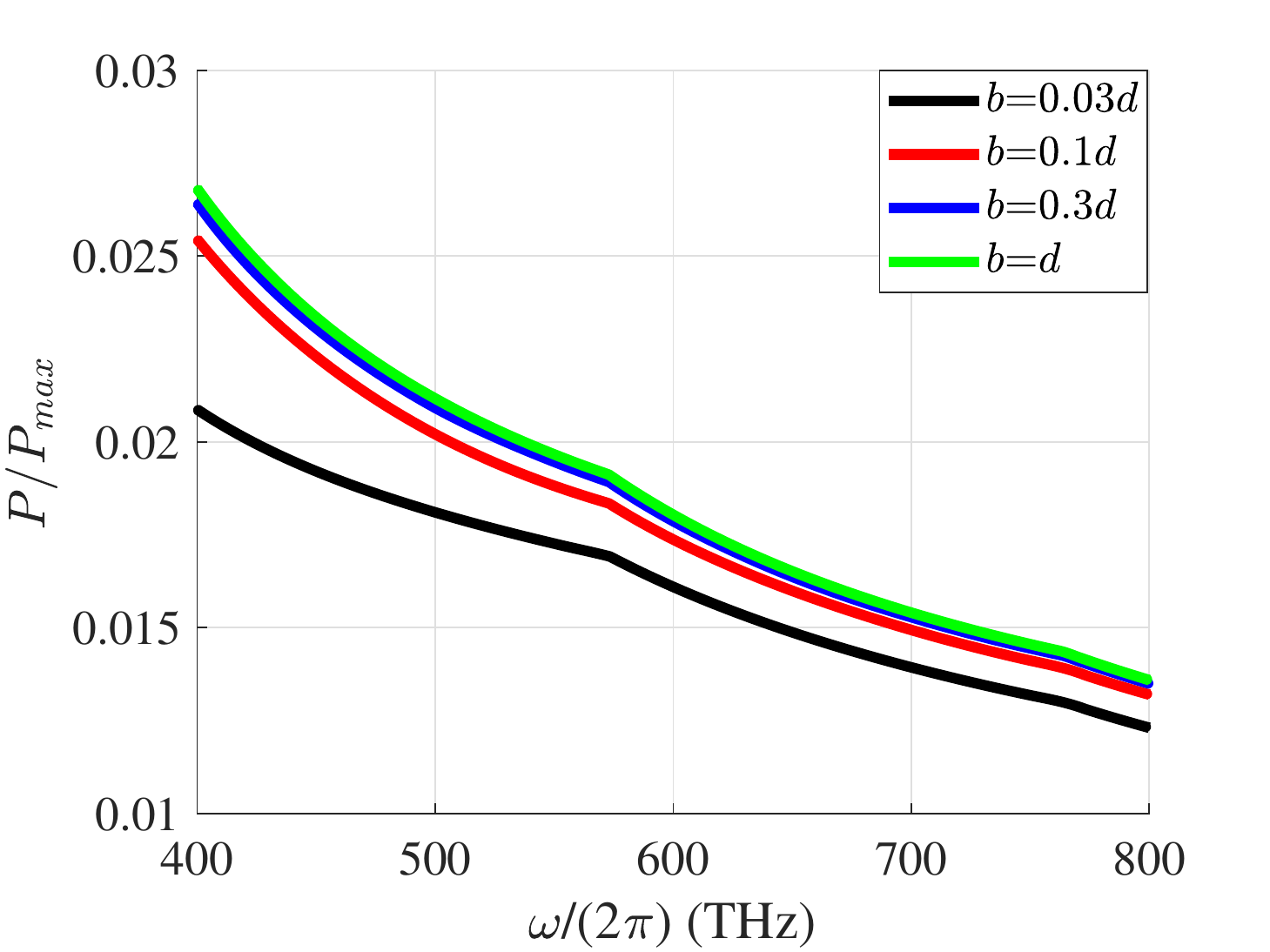}
   \label{fig:Fig4a}}
\subfigure[]{\includegraphics[width=6.0cm, draft=false]{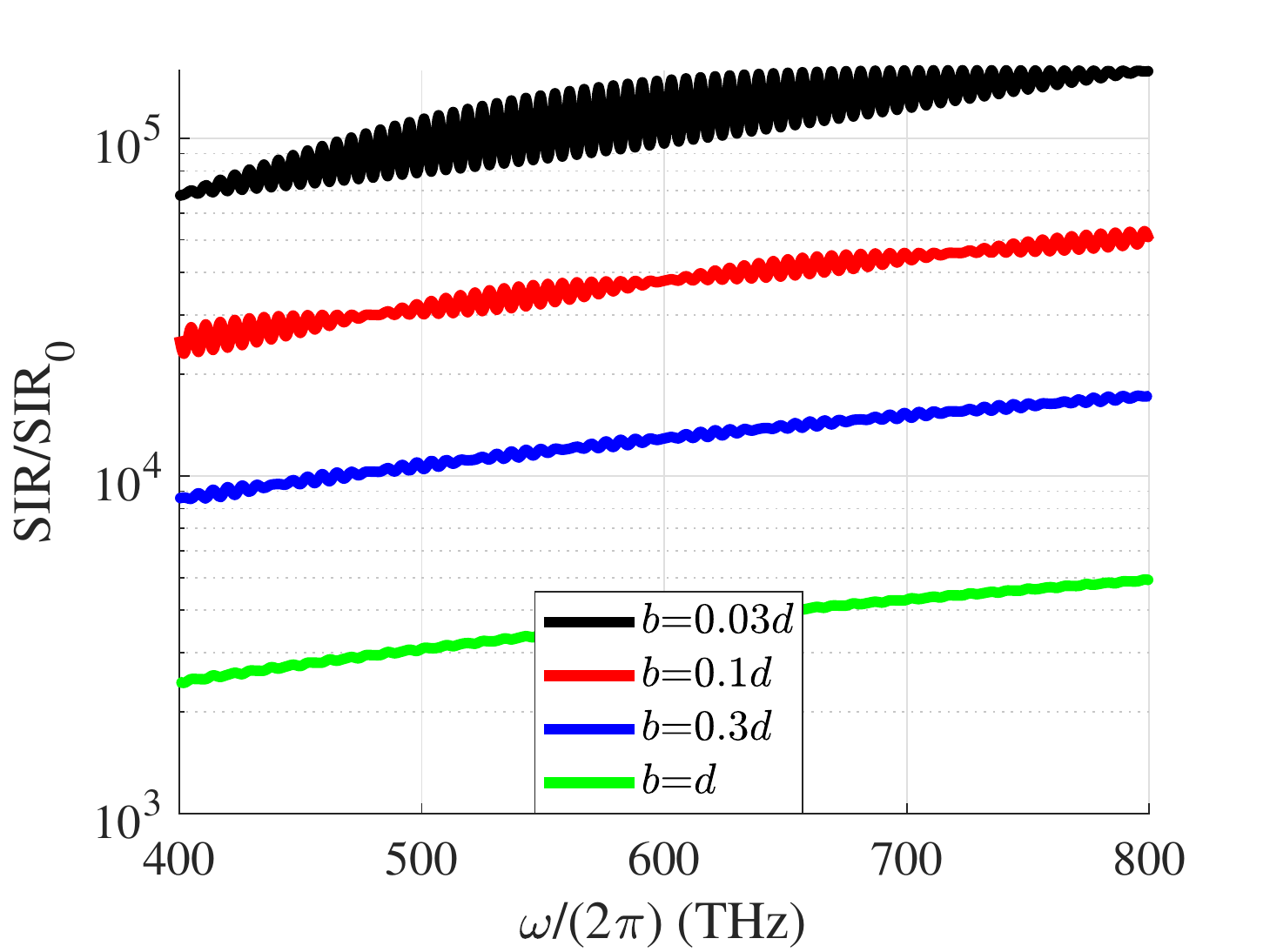}
   \label{fig:Fig4b}}
\caption{(a) Received power of the signal $P$ normalized with the total emitted power in the absence of metasurface $P_{max}$. (b) Signal-to-interference enhancement due to the metasurface ${\rm SIR/SIR_0}$. The quantities are represented as functions of the operational frequency $\w/(2\pi)$ expressed in ${\rm THz}$ for several receiving aperture sizes $b$ expressed in terms of the LED emitting aperture $d$. Plot parameters: $d=2.5~{\rm mm}$, $a=25~{\rm nm}$, $h=1000~{\rm mm}$, $M=12000$, $D=2d$.}
\label{fig:Figs4}
\end{figure}

In Fig. \ref{fig:Fig4a}, we show the variation of $P/P_{max}$ with respect to operational frequency $\w/(2\pi)$ for several sizes $b$ of the receiving aperture compared to the length $d$ of the emitting metasurface. We observe a decline in the measured quantity when higher frequencies are used due to the well-recorded increased directivity combined with the same peak power (illustrated by Fig. \ref{fig:Fig2a}), which is responsible for a similar trend in Fig. \ref{fig:Fig3a}. However, the most interesting feature of Fig. \ref{fig:Fig4a} is that the size of photodiode $b$ does not play a crucial role in the received signal; indeed, by making it $35$ times larger, one can just take a moderate average increase by less than $20\%$. This result shows how superior job our metasurface does in concentrating the vast portion of the emitted power across a small angular extent via highly directive patterns. Indeed, if $M$ is high enough, a tiny reception aperture around $\f=90^o$ is enough in capturing most of the radiated signal, as indicated in Fig. \ref{fig:Fig2b}.

In Fig. \ref{fig:Fig4b}, we show the improvement in signal-to-interference ratio ${\rm SIR/SIR_0}$ for the same setups of Fig. \ref{fig:Fig4a}. We again notice giant enhancement, up to $5$ orders of magnitude, which demonstrates the efficiency of the proposed structure. Moreover, we notice that better results are achieved with smaller $b$ meaning that ${\rm SIR_0}$ is very low (close to unity) for a tiny $b$. Indeed, when the nanoslits are not used, the power received by the two LEDs (emitting and interfering one) are similar due to the Lambertian patterns of the sources giving mild variations with respect to angle $\f$. One also observes the oscillations of the curves when sweeping $\w$ which get far more substantial (especially, if one takes into account the logarithmic scale of vertical axis) when $b$ is decreased. This behavior is anticipated since a small aperture $b$ is very sensitive to changes in the spatial distribution of the signal occurring when varying $\w$; in the limit of $b\rightarrow 0$, the VLC user records the signal and the interference at a single point which are substantially dependent on the operational frequency. Finally, a mild upward slope in ${\rm SIR/SIR_0}$ curves is noted for increasing $\w$ as also happening in Fig. \ref{fig:Fig3b} since directivity boosts that relative metric. 

By combining the findings of Figs. \ref{fig:Fig4a} and \ref{fig:Fig4b}, one infers that a smaller photodiode (of length $b$) collects slightly less signal power but achieves massive ${\rm SIR}$ enlargement. Thus, such a selection may be preferable in a realistic setup of a cluster of multiple parallel VLC links along vertical axis $y$, as long as the noise levels are kept low (a signal with less $P$ is always more vulnerable to random noise of specific spectral density). A similar trade-off is observed for different operational frequencies (different color of visible light): as $\w$ increases, $P/P_{max}$ reduces but ${\rm SIR}$ improves. One may finally advocate that the extremely high enhancement in ${\rm SIR}$ in the presence of metasurfaces, reported by Fig. \ref{fig:Fig4b} (and \ref{fig:Fig3b}), will be further (and impressively) augmented when multiple LEDs are considered; indeed, in the latter case, the interference will be almost the same with the nanoslits but much higher without them.

\subsection{Receiver Misalignment}
\begin{figure}[ht!]
\centering
\subfigure[]{\includegraphics[height=3.5cm, draft=false]{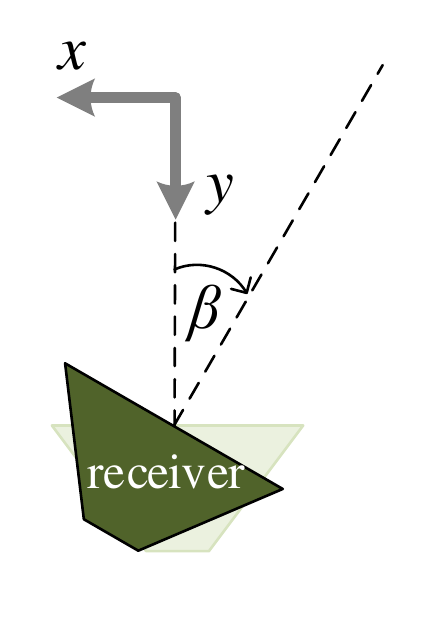}
   \label{fig:Fig45a}}
\subfigure[]{\includegraphics[height=3.5cm, draft=false]{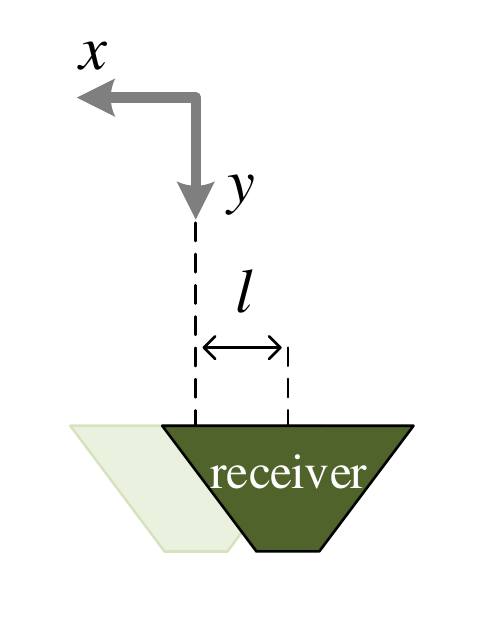}
   \label{fig:Fig45b}}
\subfigure[]{\includegraphics[height=3.5cm, draft=false]{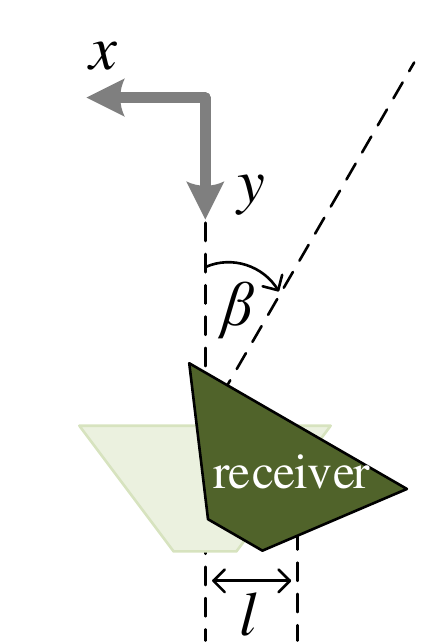}
   \label{fig:Fig45c}}
\caption{Sketches of misaligned receiver compared to its (shadow) ideal placement with reference to Fig. \ref{fig:Fig1}. (a) Imperfect angular orientation by angle $\beta$. (a) Imperfect translational position by length $l$. (c) Combined misalignment of both angle $\beta$ and position $l$.}
\label{fig:Figs45}
\end{figure}

One of the most significant defects in VLC systems is the imperfect placement of the receiver with respect to the transmitter, especially when more than one links are proposed to work in parallel like in the configuration of Fig. \ref{fig:Fig1}. The photodiode can be misplaced in terms of direction by angle $\beta$ defined in Fig. \ref{fig:Fig45a}, position by horizontal translational shift $l$ defined in Fig. \ref{fig:Fig45b} or both (Fig. \ref{fig:Fig45c}). Thus, it will be meaningful to examine the variation of the basic metrics of our design (absolute power $P/P_{max}$ and relative signal-to-interference ratio enhancement ${\rm SIR/SIR_0}$), when the link between the VLC user and the transmitter is not perfectly aligned.

\begin{figure}[ht!]
\centering
\subfigure[]{\includegraphics[width=4.2cm, draft=false]{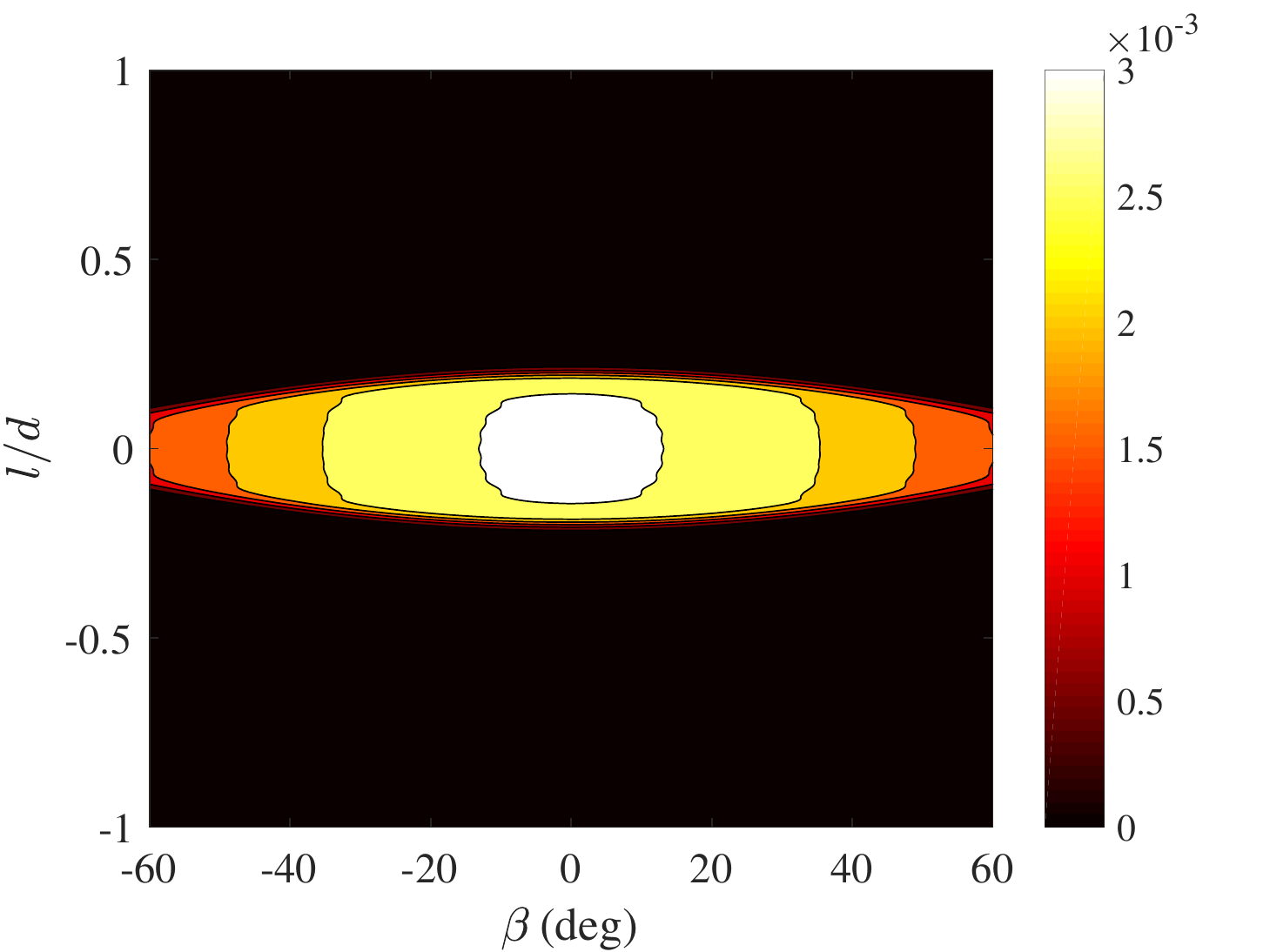}
   \label{fig:Fig5a}}
\subfigure[]{\includegraphics[width=4.2cm, draft=false]{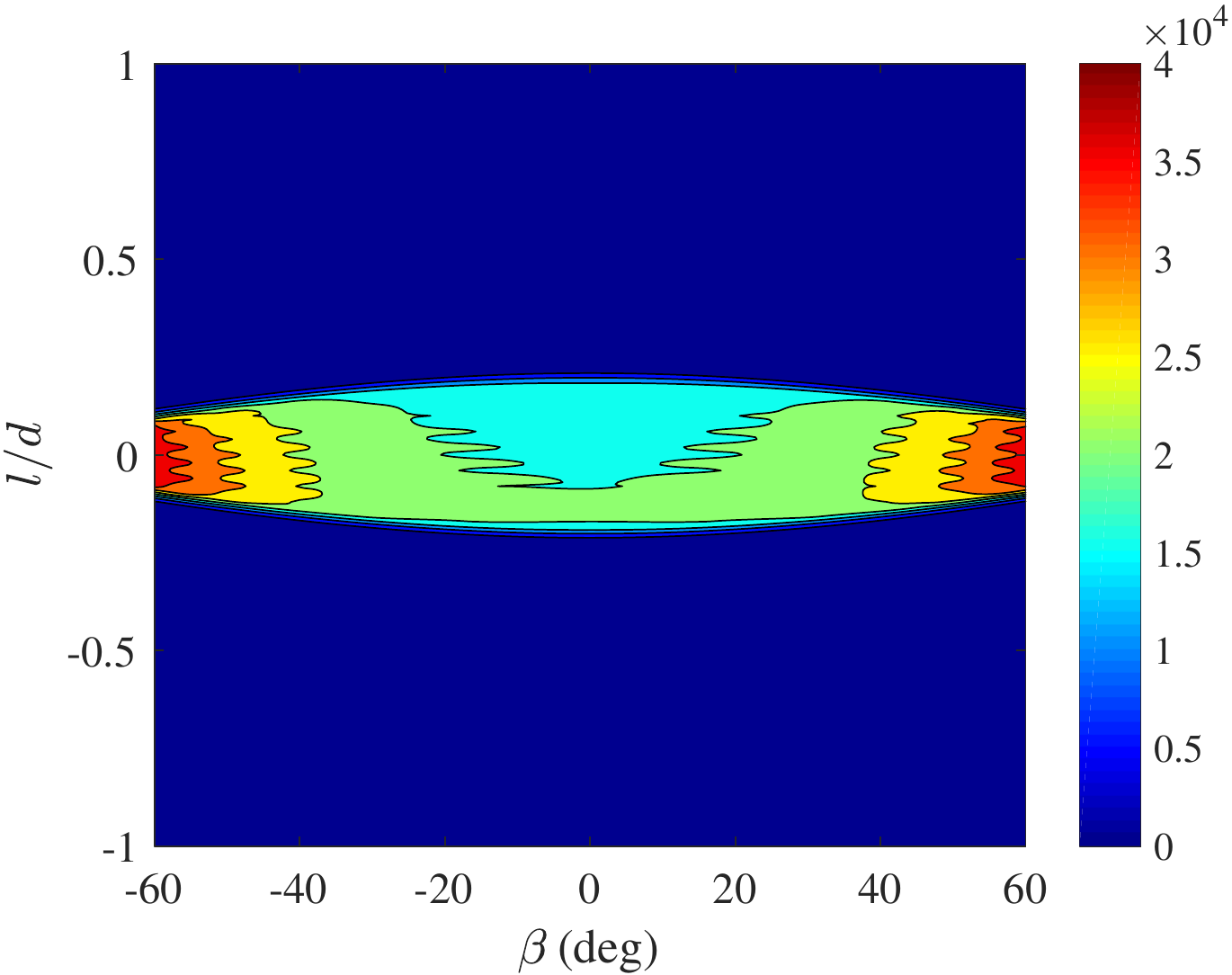}
   \label{fig:Fig5b}}\\
\subfigure[]{\includegraphics[width=4.2cm, draft=false]{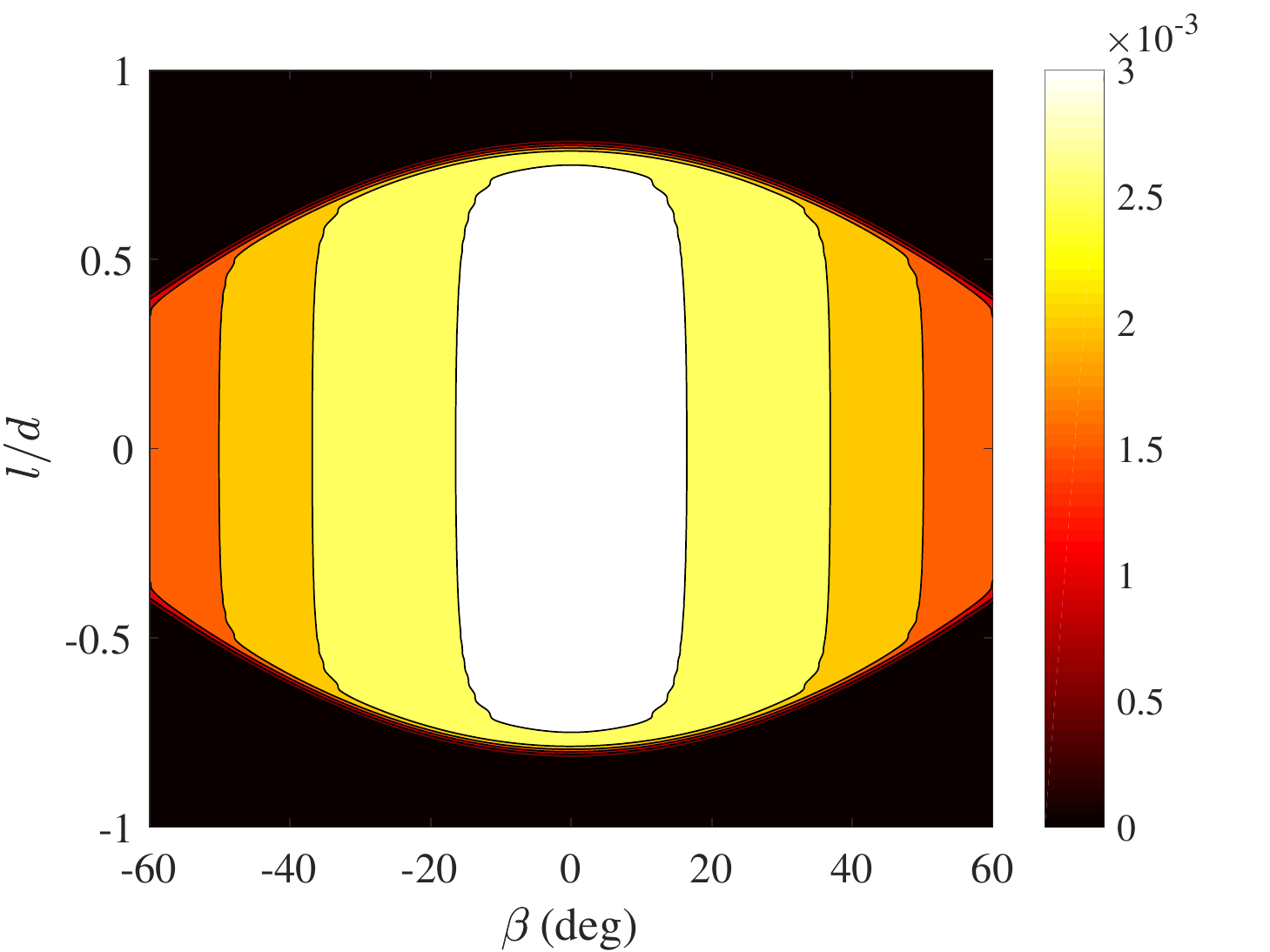}
   \label{fig:Fig5c}}
\subfigure[]{\includegraphics[width=4.2cm, draft=false]{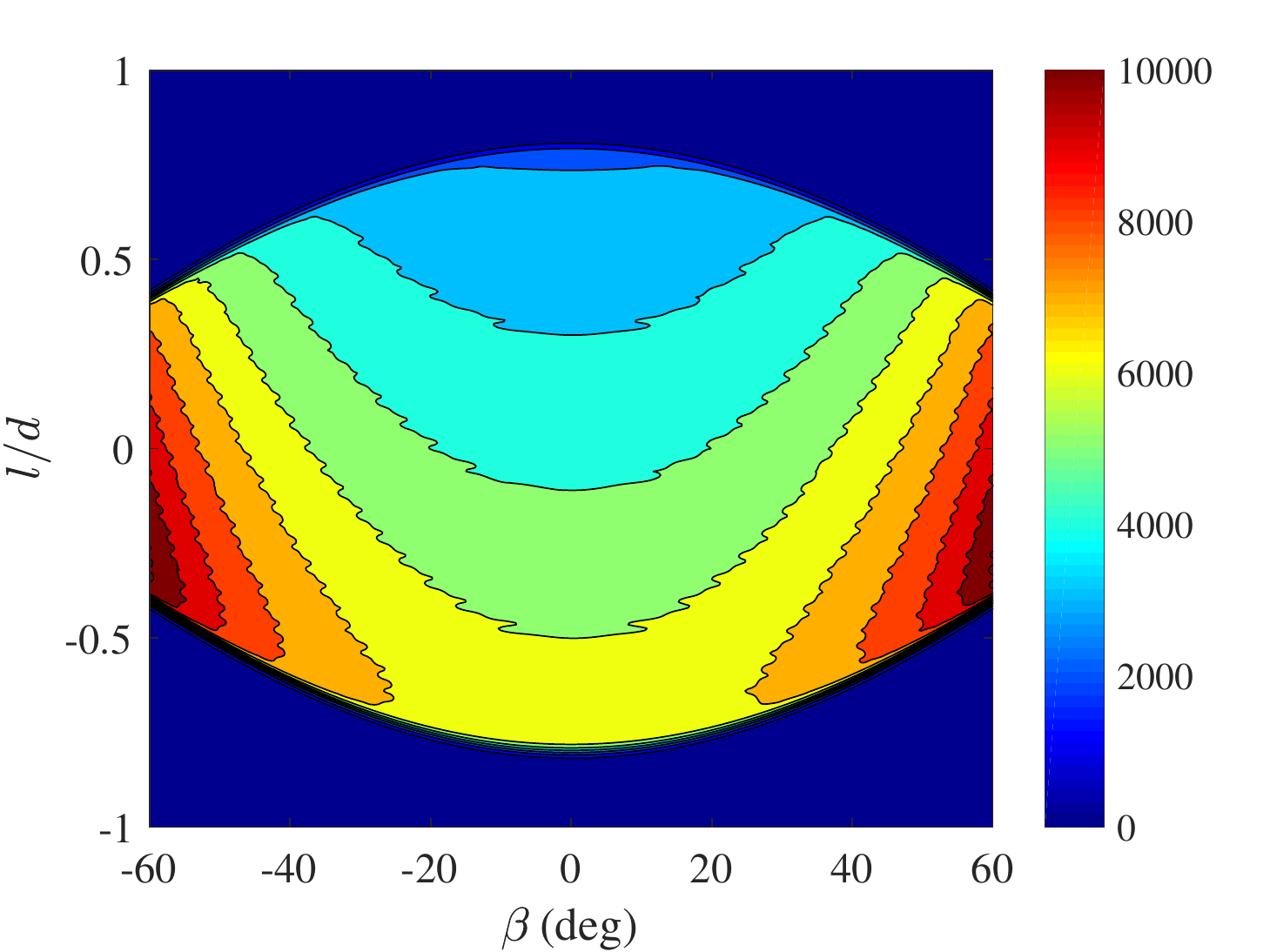}
   \label{fig:Fig5d}}
\caption{Contour plots of (a, c) $P/P_{max}$ and (b, d) ${\rm SIR/SIR_0}$ with respect to the angle $\beta$ and the position $l/d$ of a misaligned receiver. The length of the photodiode aperture is equal to: (a, b) $b=0.2d$ and (c, d) $b=0.8d$. Plot parameters: $d=2.5~{\rm mm}$, $a=25~{\rm nm}$, $h=1000~{\rm mm}$, $M=5000$, $D=2d$, $\w=(2\pi)600~{\rm THz}$.}
\label{fig:Figs5}
\end{figure}

In Fig. \ref{fig:Fig5a}, we show the quantity $P/P_{max}$ on the ``defects plane'' with the angle $\beta$ at the horizontal axis and shift $l$ divided by the fixed aperture $d$ along the vertical axis for the case of a small receiving extent $b=0.2d$. Naturally, the power gets maximized for a flawless placement of the receiver, namely at the center of the map $(\beta,l)=(0,0)$. In addition, the received signal exhibits substantial immunity with respect to rotation $\beta$ retaining half of its value even for very skewed angles of $\beta=\pm 60^o$ (for $l=0$). As far as the effect of the translational shift is concerned, it can be dramatic as long as the major lobe (like the ones of Fig. \ref{fig:Figs2}) of the emitting beam does not cross (i.e. meet spatially with) the receiver. Indeed, an approximate condition when replacing the major lobe by a single line (delta-function directivity) at direction $x=0$ reads: $-b \cos\beta<l<b \cos\beta$. Such an assumption is valid for large $M$, which is the case in the described concept with non-vanishing $P$ and fully demonstrated in the plot of Fig. \ref{fig:Fig5a}. Note that $l$ can be safely varying within this range without practically reducing the power $P$ (for fixed $\beta$). It is, therefore, inferred that the VLC link exhibits robustness with respect to horizontal position $l$ of the VLC user (represented as the non-black region of the map of Fig. \ref{fig:Fig5a}); such an insensitivity is proportional to the size of the aperture $b$ and the tilt factor $\cos\beta$. 

When it comes to the enhancement ${\rm SIR/SIR_0}$, the condition to have non-negligible values remains the same ($|l|<b\cos\beta$) and indicated in Fig. \ref{fig:Fig5b}; obviously, significant ${\rm SIR}$ cannot be achieved if the receiver does not ``see'' the major lobe of the metasurface-loaded LED. Note that, unlike Fig. \ref{fig:Fig5a}, the pattern is not symmetric neither with respect to line $\beta=0$ nor with respect to line $l=0$; such a feature is attributed to the unilateral interference of our model. To put it alternatively, if in Fig. \ref{fig:Fig1}, we regarded two interfering LEDs from both sides of the main one, the metric ${\rm SIR/SIR_0}$ would exhibit a symmetric variation on $(\beta,l/d)$ plane like $P/P_{max}$ does in Fig. \ref{fig:Fig5a}; indeed, $P/P_{max}$ is evaluated as if the basic VLC link operates alone. By inspection of Fig. \ref{fig:Fig5b}, one notices that ${\rm SIR/SIR_0}$ decreases more abruptly for $l>0$ than for $l<0$, a trend explained again by the fact that interfering signal is emanated from the one side only ($x<0$). Furthermore, the maximal values for ${\rm SIR/SIR_0}$ are recorded for large tilts $\beta$, where the interference diminishes more rapidly than the information signal $P$ (whose decline is presented in Fig. \ref{fig:Fig5a}). Once more, we notice that if the noise level is low, one can trade a small portion of its signal power $P$ for an impressive suppression of interference, namely picking a slightly tilted collector.  

In Figs. \ref{fig:Fig5c} and \ref{fig:Fig5d}, the same metrics as in Figs. \ref{fig:Fig5a} and \ref{fig:Fig5b} respectively, are plotted but for a VLC user with a lengthier aperture of $b=0.8d$. Similar conclusions are drawn with the difference that the parametric region of non-negligible output is more extended due to the size of $b$ allowing the receiver higher horizontal shifts without losing the major lobe of the directive incident beam. It should be remarked that, in this case, the variation of ${\rm SIR/SIR_0}$ is almost symmetric with respect to the line $\beta=0$, since the larger photodiode ``averages'' more efficiently the signal rendering the result practically indifferent on the sign of $\beta$. Finally, notice that the interference enhancement ${\rm SIR/SIR_0}$ is much smaller compared to that of Fig. \ref{fig:Fig5b} scenario because a longer receiver will always be more vulnerable to unwanted signals (including noise). On the contrary, the beneficial power $P/P_{max}$ possesses similar magnitudes with that of Fig. \ref{fig:Fig5a} since it is dictated mainly by the major lobe which, in turn, is solely determined by the identity characteristics of the patterned LED aperture ($a, d$). 

Note that we can evaluate the received power $P_0$ in the absence of directivity metasurfaces, with help from the Poynting theorem like in \eqref{SIRFormula} and \eqref{SIRFormula0}; indeed, the description of the receiving aperture in the Cartesian coordinate system reads: $y(x)=h-(x+l)\tan\beta$ for $|x+l|<b\cos\beta$, while the normal vector to its surface reads: $\hat{\textbf{n}}=\hat{\textbf{x}}\sin\beta+\hat{\textbf{y}}\cos\beta$. The ratio $P/P_0$ is obviously dependent on the size $b$ and in most cases can vastly surpass unity for a properly substantial number of slits $M$. In fact, the value of $P/P_0$ is only bounded by the constraint $M\ll N$ for the model of Fig. \ref{fig:Fig1} to be valid.

Apart from the interference from the neighboring LED, an important cause for the quality drop of any kind of communication is the presence of noise at the receiver (illustrated in Fig. \ref{fig:Fig1}). Needless to say that in case only one frequency $\w=\w_0$ is excited, the signal is totally protected from the local white noise. On the contrary, if the bandwidth of the signal is $2\Delta\w$ around $\w=\w_0$, the power of the noise equals to $P_n=2n\Delta\w$, where $n$ is the unilateral noise spectral density (measured in $Watt/Hertz$). In this way, the signal-to-noise ratio is defined as ${\rm SNR}=P/P_n$ in the presence and as ${\rm SNR}_0=P_0/P_n$ in the absence of the metasurface. It is obvious that the signal-to-noise ratio enhancement ${\rm SNR}/{\rm SNR}_0$ is equal to the actual power ratio $P/P_0$ and, thus, amplifiable at will via suitably large number $M\ll N$ nanoslits. On the contrary, the quantity that should be taken into account to entertain white noise influence (of spectral density $n$) on an information signal (of bandwidth $2\Delta\w$) is our absolute metric $P/P_{max}$. Consider also that $P_{max}$ is a (rather mildly varying) function of frequency $\w$, namely $P_{max}=P_{max}(\w)\sim 1/\w^2$ since it is dependent on $k_0=\w/c$ (for fixed metasurface-based LED emitter with specific $(a,d)$). Nonetheless, the bandwidth in practice is a tiny fraction of the modulating frequency (i.e. $2\Delta\w \ll \w_0$) and thus the normalized power $P/P_{max}={\rm SNR}~\w_0~\frac{(2\Delta\w/\w_0)n}{P_{max}(\w_0)}$ determines the ${\rm SNR}$ for given parameters $n$, $\Delta\w$ at a frequency $\w=\w_0$.

\section{Concluding Remarks}
\label{sec:conclusion}

The key challenge of visible-light communication systems is the strong interference between the multiple LED sources working in parallel and in close proximity each other. A simple, yet very efficient, concept based on the deployment of a primitive metasurface in the near field of the sources, that forms remarkably directive emitting beams, has been proposed in the present work. Since such a system manages to concentrate the vast portion of the incident power into a very small angular extent, it makes the communication practically immune to both inter-LED intervention and reflections from the indoor surroundings. In particular, impressive enhancement of the signal-to-interference ratio is recorded when using the suggested nanoslit screen, exhibiting significant robustness with respect to changes in the position and orientation of the receiver.

Our study, even though entertaining the case of a single interfering source, can be easily generalized to treat networks of LEDs placed arbitrarily in three-dimensional space. Importantly, the followed mathematical formulation is rigorous and valid for any operating frequency; it is, therefore, straightforward to be applied in VLC integrating systems supporting not only download at visible frequencies but also upload with millimeter waves \cite{MmWaveUpload}. Given the fact that VLC moves fast towards the direction of standardization and large-scale commercialization, our solution can ignite implementation efforts for hardware fabrication offering unprecedented levels of reliability at extremely high data rates.


\begin{thebibliography}{00}

\bibitem{FirstPaper}
T. Komine and M. Nakagawa,
``Fundamental Analysis for Visible-Light Communication System using LED Lights,'' 
\emph{IEEE Transactions on Consumer Electronics}, vol. 50, no. 1, pp. 100-107, 2004.

\bibitem{BookDimitrovHaas}
S. Dimitrov and H. Haas, 
\emph{Principles of LED Light Communications: Towards Networked Li-Fi}, 
Cambridge, UK: Cambridge University Press, 2015.

\bibitem{BookArnon} 
S. Arnon, 
\emph{Visible Light Communication}, 
Cambridge, UK: Cambridge University Press, 2015.

\bibitem{VLCStateOfArt} 
D. Karunatilaka, F. Zafar, V. Kalavally, and R. Parthiban,
``LED Based Indoor Visible Light Communications: State of the Art,''
\emph{IEEE Communications Surveys and Tutorials}, vol. 17, no. 3, pp. 1649-1678, 2015.

\bibitem{VLCChallenges} 
P. H. Pathak, X. Feng, P. Hu, and P. Mohapatra,
``Visible Light Communication, Networking, and Sensing: A Survey, Potential and Challenges,''
\emph{IEEE Communications Surveys and Tutorials}, vol. 17, no. 4, pp. 2047-2077, 2015.

\bibitem{HighSpeedVLC}
L. Grobe, A. Paraskevopoulos, J. Hilt, D. Schulz, F. Lassak, F. Hartlieb, C. Kottke, V. Jungnickel, and K.-D. Langer,
``High-Speed Visible Light Communication Systems,''
\emph{IEEE Communications Magazine}, vol. 51, no. 12, pp. 60-66, 2013.

\bibitem{VLCPoint2Point}
H. Burchardt, N. Serafimovski, D. Tsonev, S. Videv, and H. Haas,
``VLC: Beyond Point-to-Point Communication,''
\emph{IEEE Communications Magazine}, vol. 52, no. 7, pp.  98-105, 2014.

\bibitem{VLCSecurity}
L. Yin and H. Haas,
``Physical-Layer Security in Multiuser Visible Light Communication Networks,''
\emph{IEEE Journal on Selected Areas in Communications}, vol. 36, no. 1, pp. 162-174, 2018.

\bibitem{VLCMarket} 
A. Jovicic, J. Li, and T. Richardson,
``Visible Light Communication: Opportunities, Challenges and the Path to Market,''
\emph{IEEE Communications Magazine}, vol. 51, no. 12, pp. 26-32, 2013.

\bibitem{PureLiFi} 
Pure Li-Fi company website: \hyperlink{https://purelifi.com}{https://purelifi.com}.


\bibitem{InterferenceManagement} 
M. Kashef, M. Abdallah, K. Qaraqe, H. Haas, and M. Uysal,
``Coordinated Interference Management for Visible Light Communication Systems,''
\emph{IEEE/OSA Journal of Optical Communications and Networking}, vol. 7, no. 11, pp. 1098-1108, 2015.

\bibitem{DemonstrationBiDirectional}
Y. F. Liu, C. H. Yeh, C. W. Chow, Y. Liu, Y. L. Liu, and H. K. Tsang,
``Demonstration of bi-directional LED visible light communication using TDD traffic with mitigation of reflection interference,''
\emph{Optics Express}, vol. 20, no. 21, pp. 23019-23024, 2012.

\bibitem{InterferenceRejection}
C.-C. Chang, Y.-J. Su, U. Kurokawa, and B. I. Choi,
``Interference Rejection Using Filter-Based Sensor Array in VLC Systems,''
\emph{IEEE Sensors Journal}, vol. 12, no. 5, pp. 1025-1032, 2012.

\bibitem{IndoorChannel}
K. Lee, H. Park, and J. R. Barry,
``Indoor Channel Characteristics for Visible Light Communications,''
\emph{IEEE Communications Letters}, vol. 15, no. 2, pp. 217-219, 2011.

\bibitem{ChannelModeling} 
F. Miramirkhani and M. Uysal,
``Channel Modeling and Characterization for Visible Light Communications,''
\emph{IEEE Photonics Journal}, vol. 7, no. 6, pp. 7905616, 2015.


\bibitem{SmithReview}
C. L. Holloway, E. F. Kuester, , J. A. Gordon, J. O'Hara, J. Booth, and D. R. Smith,
``An Overview of the Theory and Applications of Metasurfaces: The Two-Dimensional Equivalents of Metamaterials,''
\textit{IEEE Antennas and Propagation Magazine}, vol. 54, no. 2,  pp.~10-35, 2012.

\bibitem{Generation3DCusp}
W. Liu, Y. Zhang, J. Gao, and X. Yang,
``Generation of three-dimensional optical cusp beams with ultrathin metasurfaces,''
\emph{Scientific Reports}, vol. 8, pp. 9493, 2018.

\bibitem{OptoVLSIBeamformer}
F. Xiao, K. Alameh, and Y. T. Lee,
``Tunable multi-wavelength fiber lasers based on an Opto-VLSI processor and optical amplifiers,''
\emph{Optics Express}, vol. 17, pp. 23123, 2009.

\bibitem{OrbitalAngularMomentum}
M. L. N. Chen, L. J. Jiang, and W. E. I. Sha,
``Orbital Angular Momentum Generation and Detection by Geometric-Phase Based Metasurfaces,''
\emph{Applied Sciences}, vol. 8, pp. 362, 2018.

\bibitem{AnalyticTheory}
J. W. Yoon, M. J. Jung, S. H. Song, and R. Magnusson,
``Analytic Theory of the Resonance Properties of Metallic Nanoslit Arrays,''
\emph{IEEE Journal of Quantum Electronics}, vol. 48, no. 7, pp. 852-861, 2012.

\bibitem{OpticalSwitching}
M. Jarrahi, R. F. W. Pease, D. A. B. Miller, and T. H. Lee,
``Optical switching based on high-speed phased array optical beam steering,''
\emph{Applied Physics Letters}, vol. 92, pp. 014106, 2008.

\bibitem{ActiveControl}
S. Jia, X. Wang, Y. Wu, M. Xiao, P. Fan, and Z. Wang,
``Active control of beams by metallic nanoslit array lens with movable dielectric substrate,''
\emph{Applied Physics Express}, vol. 8, no. 2, pp. 062001, 2015.

\bibitem{NanoslitArrays}
K.-L. Lee, J.-B. Huang, J.-W. Chang, S.-H. Wu, and P.-K. Wei,
``Ultrasensitive Biosensors Using Enhanced Fano Resonances in Capped Gold Nanoslit Arrays,''
\emph{Scientific Reports}, vol. 5, pp. 8547, 2015.

\bibitem{AntennaBalanis} 
C. A. Balanis, 
\emph{Antenna Theory}, 
New York, New York, USA: John Wiley \& Sons, 1997.

\bibitem{CoordinatedBeamforming}
H. Ma, A. Mostafa, L. Lampe, and S. Hranilovic,
``Coordinated Beamforming for Downlink Visible Light Communication Networks,''
\emph{IEEE Transactions on Communications}, vol. 66, no. 8, pp. 3571-3582, 2018.

\bibitem{ImpactLEDTransmitters}
Y. Wang, M. Chen, J.-Y. Wang, J. Shi, Z. Yang, Y. Pan, and R. Guan,
``Impact of LED transmitters’ radiation pattern on received power distribution in a generalized indoor VLC system,''
\emph{Optics Express}, vol. 25, no. 19, pp. 22805-22819, 2017.

\bibitem{CylWaves} 
C. A. Valagiannopoulos,
``On smoothening the singular field developed in the vicinity of metallic edges,''
\emph{International Journal of Applied Electromagnetics and Mechanics}, vol. 31, no. 2, pp. 67-77, 2009.

\bibitem{Abramowitz} 
M. Abramowitz and I. Stegun, 
\emph{Handbook of Mathematical Functions},
New York, New York, USA: Dover Publications, 1964.

\bibitem{Pulses}
J. G. Proakis and M. Salehi, 
\emph{Communication Systems Engineering},
Upper Saddle River, New Jersey, USA: Prentice Hall, 2002.

\bibitem{FluxIntensity} 
A. E. F. Taylor, 
\emph{Illumination Fundamentals}, 
New York, New York, USA: Lighting Research Center at Rensselaer Polytechnic Institute, 2000.

\bibitem{LambertianPattern} 
I. Moreno and C.-C. Sun,
``Modeling the radiation pattern of LEDs,''
\emph{Optics Express}, vol. 16, no. 3, pp. 1808-1819, 2008.

\bibitem{Judicious}
C. A. Valagiannopoulos and V. Kovanis,
``Judicious distribution of laser emitters to shape the desired far-field patterns,''
\emph{Physical Review A}, vol. 95, pp. 063806, 2017.

\bibitem{MmWaveUpload}
L. Feng, H. Yang, R. Q. Hu and J. Wang,
``MmWave and VLC-Based Indoor Channel Models in 5G Wireless Networks,''
\emph{IEEE Wireless Communications}, vol. 25, no. 5, pp. 70-77, 2018.

\end{thebibliography}
\end{document}